\documentclass[notitlepage,12pt,tightenlines,eqsecnum,nofootinbib,floatfix,prd,letterpaper,longbibliography,sort&compress,sort,numbers]{revtex4-1}

\usepackage{hyperref}

\usepackage{wrapfig}
\usepackage{geometry}
\usepackage{color,xcolor}
\usepackage[most]{tcolorbox}
\usepackage[space]{grffile}
\usepackage{latexsym}
\usepackage{textcomp}
\usepackage{longtable}
\usepackage{multirow,booktabs}
\usepackage{amsfonts,amsmath,amssymb}
\usepackage[english]{babel}
\usepackage{graphicx}
\usepackage[space]{grffile}
\usepackage{url}
\usepackage[utf8]{inputenc}
\usepackage{longtable}
\usepackage{fancyhdr}
\usepackage{hyperref}
\hypersetup{
   colorlinks,%
   citecolor=blue,%
   filecolor=black,%
   linkcolor=blue,%
   urlcolor=blue
}

\definecolor{arylideyellow}{rgb}{0.91, 0.84, 0.42}
\definecolor{corn}{rgb}{0.98, 0.93, 0.36}
\definecolor{pastelyellow}{rgb}{0.99, 0.99, 0.59}
\newcommand*{\myfont}{\fontfamily{pag}\selectfont}

\DeclareTextFontCommand{\textmyfont}{\myfont}

\begin{document}
\setcounter{page}{0}

\begin{titlepage}

  \centering
  \vspace*{-1.25in}
  \hspace*{-1.04in}
  \includegraphics[width=8.5in]{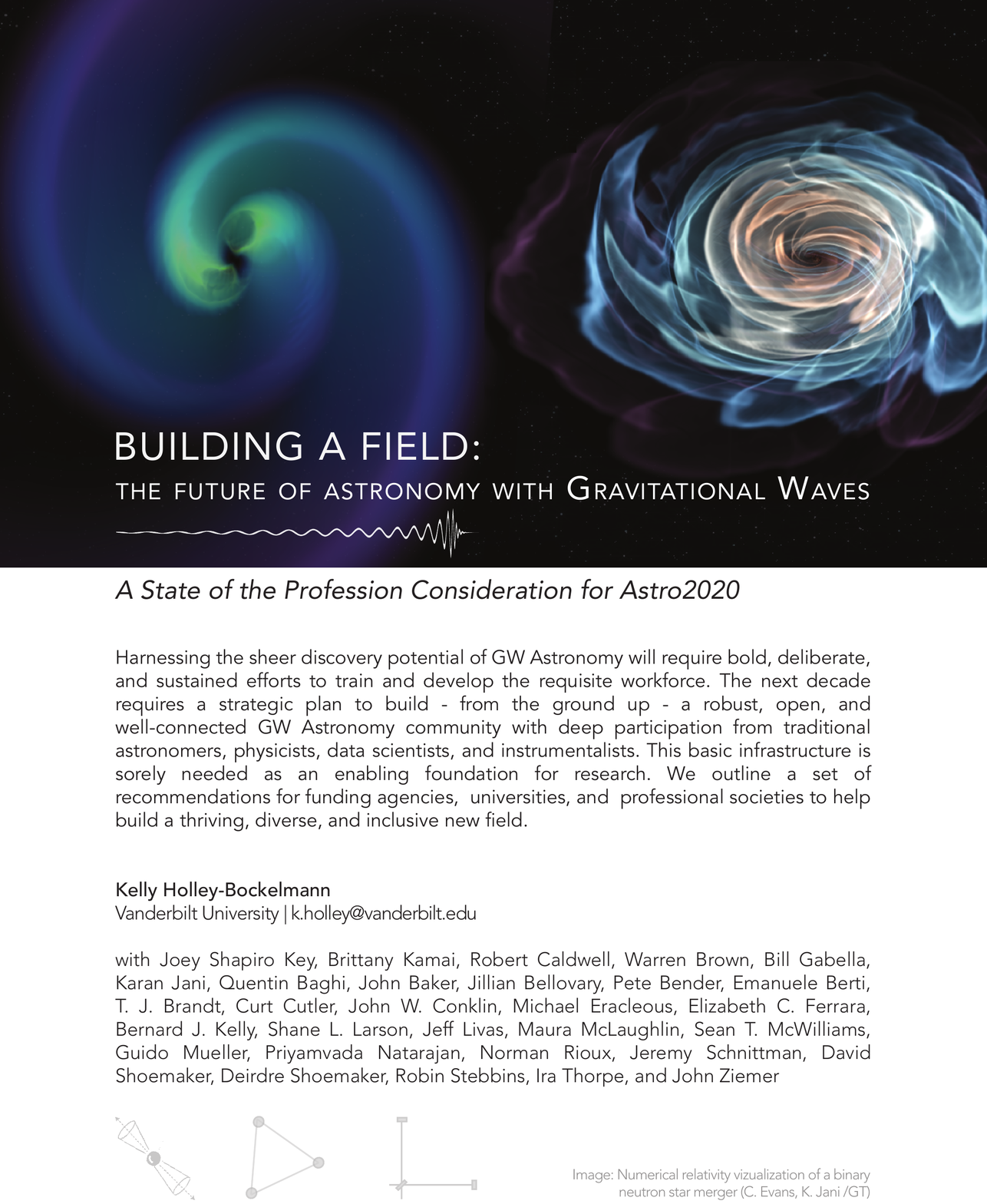}

\end{titlepage}



\date{\today}

\vspace{-15pt}
\begin{abstract}

The emerging field of gravitational wave astronomy presents a unique opportunity to build and support a scientific community to maximize gravitational wave science. Strategic efforts are currently needed in order to establish the infrastructure, collaborations, and coordination required to realize the full discovery potential of this new window on the Universe.


\end{abstract}







\begin{tcolorbox}[colback=pastelyellow,
                  colframe=black,
                  width=17cm,
                  arc=3mm, auto outer arc,
                 ]
 
\bigskip
\begin{center} {\bf \Large{ \textmyfont{Executive Summary}}} \end{center}
\medskip
{ 
Astronomy is an unusual science in that it is nearly impossible to do direct experiments; all that we know comes from observing the Universe and interpreting the messages we intercept on Earth. Until 2015, astronomers could only access light, and to a lesser degree, neutrinos and cosmic rays, to understand the distant Universe. These messengers revealed that we live in an active, expanding, and accelerating Universe -- full of exoplanets, stellar explosions, billions of other galaxies, plus dark matter and dark energy pervading everything. As technology advanced, we could observe the Universe in different wavelengths of light, and with each new electromagnetic waveband, our understanding of the Universe changed dramatically. Just four years ago, humanity first observed gravitational waves, ripples in spacetime caused by the motion of massive systems like binary black holes. Astronomers now have a window onto the Universe that was inaccessible before, a window peering into spacetime itself and carrying the message of masses in motion. 
 


\medskip 




\medskip 





 
 Gravitational Wave (GW) Astronomy has arrived, and the need for building expertise is urgent. One indicator of that rising urgency is the number of advertised positions in GW Astronomy. Data taken from the AAS Job Register shows a 50-fold increase in advertisements for faculty positions in GW and Multimessenger Astronomy from 2014 to 2018, while postdoc positions in these fields increased by two orders of magnitude. We are at a time when demand outpaces supply; with a current US and European cohort of faculty job-seekers with gravitational wave expertise numbering roughly a dozen, recent PhDs are being recruited directly into faculty positions and there have been several faculty searches unfilled. We expect this demand to continue across the entire gravitational wave spectrum as more detections and new observatories come online.
 



 \medskip
 
 It is clear that  {\bf harnessing the sheer discovery potential of GW Astronomy will require bold, deliberate, and sustained efforts to train and develop the requisite workforce. The next decade requires a strategic plan to build -- from the ground up -- a robust, open, and well-connected GW Astronomy community} with deep participation from traditional astronomers, physicists, data scientists, and instrumentalists. This basic infrastructure is sorely needed as an enabling foundation for research. We outline a set of recommendations for funding agencies,  universities, and  professional societies to help build a thriving, diverse, and inclusive new field.
 \bigskip
 
}
 \end{tcolorbox}
 
\eject

\begin{figure}
\begin{center}
\vspace{-60pt}
\includegraphics[width=12cm,angle=270]{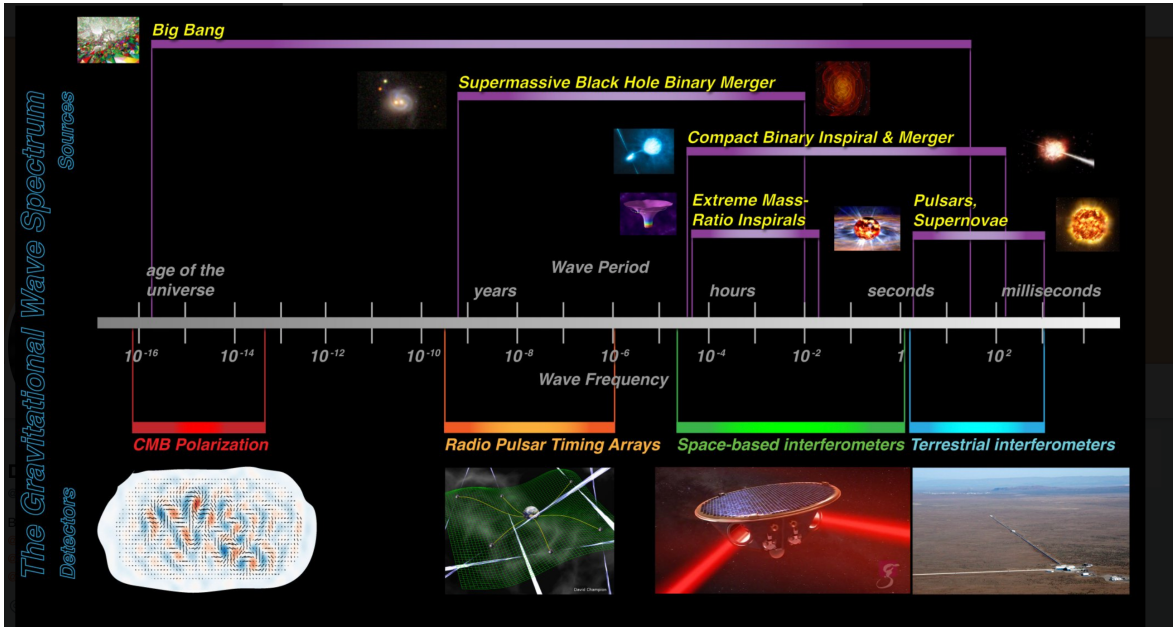}
\vspace{-50pt}
\caption{The GW spectrum spans from the lowest frequencies, oscillating at the age of the Universe (left), to the highest frequencies, rapidly spinning objects (right). The spectrum is produced by a variety of different phenomena spanning the entire history of the Universe, which requires a variety of observation techniques. (Credits: K.~Jani, Bicep-Keck \cite{bicep}, NANOGrav \cite{nanograv}, LISA \cite{lisa}, LIGO \cite{aligo}}.
\vspace{-30pt}
\label{fig:spectrum}
\end{center}
\end{figure}

\smallskip 
\noindent{\large{\textmyfont{The Launch of Gravitational Wave Astronomy}}} 

To move from inception to the first direct observation of gravitational waves has taken over 50 years of theoretical and experimental efforts. The explosion of detections has firmly launched this observational discipline into the landscape of 21st Century astronomy. GW observatories such as LIGO (Laser Interferometer Gravitational-wave Observatory) \cite{aligo, Aasi2015} and LISA (Laser Interferometer Space Antenna) \cite{lisa, L3proposal} are facility-class endeavors that require hundreds of scientists and engineers collaborating with hundreds of other skilled professionals to design, construct, and operate this new generation of observational tools, much like the flagship surveys and missions such as LSST \cite{LSST, ivezic2008lsst} and JWST \cite{jwst, Gardner2006}.

GW Astronomy is emerging almost simultaneously across four distinct parts of the GW spectrum, which is unlike traditional electromagnetic (EM) astronomy that benefitted from optical data for millennia and  developed radio and X-ray observatories over the course of the last century. As illustrated in Figure 1, the GW spectrum includes: the ultra-low frequency band, covered by cosmic microwave background polarization experiments like CMB-S4 \cite{cmbs4, Abazajian2015}; the very-low frequency band, covered by Pulsar Timing Arrays (PTAs) like NANOGrav \cite{nanograv, McLaughlin2013}; the low-frequency band, covered by LISA \cite{lisa, L3proposal}; and the high frequency band, covered by ground-based facilities like LIGO in the US, Virgo in Europe \cite{virgo, Acernese2015}, and KAGRA in Japan \cite{kagra, kagra2}.  These four gravitational wave bands have key differences in the underlying astrophysics and experimental techniques. Yet, they share certain commonalities: some data analysis techniques are extensible between the bands, and computational and data storage needs are comparable. 





GW Astronomy requires a remarkable breadth of expertise -- practitioners from nearly all sub-disciplines of physics and astronomy, as well as engineering, computer science, data science, and statistics are and will be needed to bring the field to fruition and derive the maximum scientific benefit from our GW investments. Moreover, since most astrophysics research areas will benefit from GW data to accelerate their own disciplines, ensuring access and usability of that data is essential to anticipate the rapid pace of discovery. 

\vspace{10pt} 
\noindent{\large \textmyfont{The Landscape of Future Gravitational Wave Facilities}}

Currently, the second generation ground-based facilities, Advanced LIGO and Advanced Virgo, are building a catalog of compact binary mergers, as well as piloting  partnerships with EM observers on multi-messenger GW and EM detections. With ever-enhanced  sensitivity and with new detectors coming online (e.g., KAGRA and  LIGO-India \cite{iligoindia, indigo2011}), the forecast through the next decade includes thousands of mergers, yielding estimates for the merger-rate density and mass function of local binary neutron stars and stellar-origin black holes \cite{Kalogera:2019sui,Kalogera:2019bdd}. The planned observatories, however, span the entire GW spectrum and promise a spectacular range of new science that touches on nearly every astronomical subfield.

\noindent {\bf High-frequency band - Ground-based:} 
The 3rd generation of ground-based detectors, slated for the mid-to-late 2030s will feature new facilities and an order-of-magnitude leap in sensitivity that will reach  all of the stellar-mass binaries in the Universe, leading to millions of detections per year\citep{gwicwp}. This will enable population and large-scale structure studies; the tension in the Hubble constant may be resolved and studied at different epochs. The much higher signal-to-noise detections for nearby sources will enable very sensitive tests of General Relativity, and coalescences involving matter, e.g., neutron stars, will offer information to help resolve the equation of state. 

\noindent {\bf Low-frequency band - Space-based:}  Accessing the most source-rich gravitational waveband requires a space-based detector. The astounding success of the LISA Pathfinder \cite{LISAPathfinder} mission demonstrated most of the technical capabilities for a space-based GW observatory and laid the groundwork for the European Space Agency (ESA) to launch LISA in 2034 \cite{lisa}. NASA is formalizing a partnership with ESA to participate in this revolutionary mission. LISA is designed to detect the merger of supermassive black holes at the centers of galaxies all the way to $z=20$ \cite{smbhwp}, tens of thousands of binary stellar remnants throughout the galaxy \cite{gbwp}, stars falling into massive black holes in galactic centers out to $z\sim1$ \cite{emriwp}, and exotic types of dark matter\cite{cosmowp} (see Figure 2). However, perhaps the most exciting part of the LISA mission will be its potential for discovering entirely new physical phenomena \cite{grwp, discoverywp}. 

\noindent {\bf Very-low frequency - Pulsars:} Pulsar timing arrays are on the brink of dramatically expanding the GW astronomy spectrum with a detection of the stochastic background from  supermassive black hole binaries at the cores of galaxy mergers with orbital periods of years to decades. For the first time, PTA upper limits on this background are constraining the astrophysics of galaxy mergers and black hole-bulge mass relations \cite{nanogravSB}. With the next several years, a detection of this background will offer unique insights into the process of galaxy mergers and  the next decade will see the first individual supermassive black hole binary detections, allowing joint GW/EM observations of galaxy mergers.

\noindent {\bf Ultra-low frequency band - Cosmic Microwave Background:} One of the most ambitious goals of current cosmological research is to measure gravitational waves from the very first moments of the Big Bang. Single-quantum fluctuations of gravity produce relic tensor perturbations in space-time curvature that are stretched during cosmic inflation to the largest scales. If inflation expands rapidly enough, they could appear  today as a measurable distinctive {\it B mode} pattern of anisotropic polarization in the cosmic background radiation, and CMBS4 is one of the leading next-generation ground-based cosmic microwave background polarization projects \cite{cmbs4}.

\begin{figure}
\centering
\includegraphics[width=12cm]{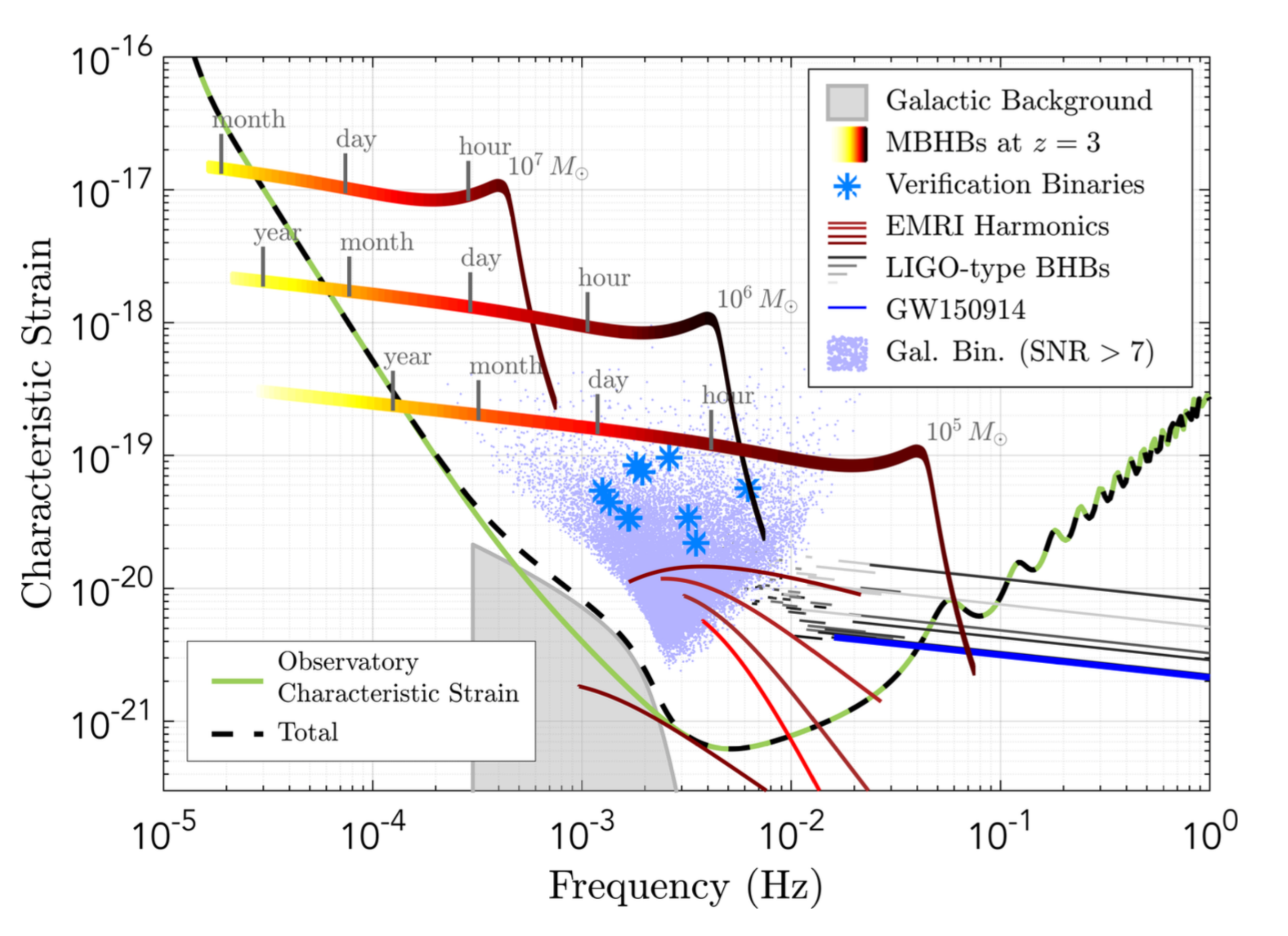}
\vspace{-15pt}
\caption{Examples of known GW sources in the LISA band, credit: LISA proposal in response to the ESA call for L3 mission concepts \cite{L3proposal}.}
\label{fig:sources}
\end{figure}

\vspace{10pt} 
\noindent{\large \textmyfont{The Case for Growing a Gravitational Wave Astronomy Community}}

With these new capabilities, GW observations can expect to undergo a transition not unlike that of exoplanets -- evolving from a state in which a single detection is novel and important -- to a data-rich state that elucidates deeper astronomical questions. And as in the case of exoplanets, in which it is becoming clear that progress requires crossing the historical divides between planetary, solar, and stellar astrophysics, the expertise needed to build GW Astronomy is housed in very different silos.

GW Astronomy will stall if we maintain the status quo. We note that nearly all GW experts received PhDs in Physics, with little or no formal training in traditional astronomy and limited professional interactions with astronomers. Gravitational waves are housed in the Division of Gravitational Physics (DGRAV) of the American Physical Society (APS), and therefore the best-attended conferences by the GW community are the April APS meeting and the Amaldi meeting, both of which are venues lacking astronomers. Likewise, the core curriculum of most astronomy PhD programs do not include general relativity, GWs, or multimessenger astronomy.



GW Astronomy promises paradigm shifts in the 2020s and 2030s, and to meet these promises, we need to shift our approach to deliberately build both broader and deeper expertise in GW Astronomy, expand multimessenger astrophysics to include GWs, enable innovation in GW data analysis in the multimessenger context, and expand GW technology development, all while inculcating an inclusive and welcoming culture to produce the best science.

\vspace{10pt} 
\noindent{\large \textmyfont{(Re)building Expertise in Gravitational Wave Astronomy}}

The 2016 mid-decadal assessment by the National Academy of Sciences criticized the loss of expertise in GW Astronomy and advised a complete turnaround in financial support to make up for lost ground \cite{Midterm}. In addition to funding, though, we need innovative solutions to foster crucial connections between researchers with GW expertise and professionally-trained astronomers. Multi-band partnerships between  LIGO \cite{aligo}, LISA \cite{lisa}, NANOGrav \cite{nanograv}, and CMB-S4 \cite{cmbs4} are also important to accelerate and improve data analysis methods and tools, coordination of multimessenger observations, and technology development to ultimately enhance the science. Efforts to build the GW Astronomy community should begin immediately, including the following recommended initiatives:

\begin{itemize}
\setlength\itemsep{0em}
\item Federally-funded {\bf annual summer bootcamps} for postdocs and faculty on GW Astronomy to equip {\it senior-level}  scientists to train future GW community.
\item Associated {\bf workshops for graduate students and undergraduate students}.
\item Support for scientists who  
contribute {\bf GW workshop materials, virtual courses, and public codes} to build and engage the professional community (see the \href{https://www.gw-openscience.org/about/}{Gravitational Wave Open Science Center (GWOSC)} \cite{GWOSC} and \href{https://www.gw-openscience.org/static/workshop1/course.html}{LIGO Open Data Workshop} \cite{LIGOopen}) 
\item{Federal support for the creation of {\bf faculty lines and cluster hires in GW Astronomy}, patterned after the NSF Faculty Development in Solar and Space Physics Program \href{https://www.nsf.gov/pubs/2014/nsf14506/nsf14506.htm}{FDSS}, would jump-start the GW scientific pipeline.} 
\item{{\bf Adapting graduate programs} to incorporate GW training in formal Astronomy PhD coursework and formal astronomy training for PhDs in GW astronomy. }
\item{Robust {\bf cross-agency and cross-divisional research funding} for GW Astronomy provides a coordinated mechanism to foster science between disciplines 
(see the Nexus for Exoplanet Systems Science \href{https://nexss.info/} {NExSS} \cite{NExSS})}
\item A {\bf Faculty and Student Team (FAST)} program for astronomers and their students to ease the barrier to entry into the existing GW community through long-term research partnerships with GW experts. The \href{https://www.sdss.org/education/faculty-and-student-team-fast-initiative/}{SDSS FAST initiative} is an example of this type of program: In the case of SDSS FAST, faculty-student teams at minority institutions receive hands-on training and conduct cutting-edge research using SDSS data with established SDSS collaborators, allowing the faculty to develop expertise in a new field and provide high-quality research opportunities to underrepresented students. Building capacity at the faculty level magnifies the effort as they “pay-it-forward" to many students in the long term.\cite{SDSSfast}. 
\item {\bf Graduate Student GW Traineeships} to provide a cost-effective way of building capacity within an existing GW Astronomy group or to support establishing new one; unlike individual investigator grants, which are designed to fund a particular research project, this program could be deployed more fluidly to train graduate students in the field, and would feature extensive GW workshops, professional development, and research mentoring by experts in the field.
\item A {\bf GW Postdoctoral Fellows} program for new PhDs interested in transitioning into the GW field could take advantage of the same workshop, mentoring and training infrastructure to build additional capacity.
\item {\bf GW Ambassadors} recruited and trained to communicate with the greater astronomical community about GW Astronomy and the capabilities of future missions. GW Ambassadors would be an effective way to reach primarily undergraduate and  minority-serving institutions, as well as community and tribal colleges.
\item Federal support for {\bf GW Data Challenge} efforts, enabling development of data analysis capabilities, exploring multi-band synergies, and reducing the barrier of entry to the field.
\item A {\bf GW Guest Investigator Program} to facilitate research in gravitational wave theory, experiment, and data analysis in any part of the spectrum. 
\item Development of {\bf Observer Tools}, enabling astronomers who are not GW specialists to easily evaluate, e.g.,  LISA mission capabilities.
\item Early creation of a {\bf LISA Science Center} to catalyze the growth of the space-based GW Astronomy community (see below). 
\end{itemize}



\vspace{10pt} 
\noindent{\large \textmyfont{Multimessenger Astronomy with Gravitational Waves}}

Bringing together data from different messengers  -- light, gravitational waves, and particles -- is already revolutionizing astronomy.  Our first joint EM and GW source made this abundantly clear: nearly one-third of astronomers worldwide observed the double neutron star merger, 
GW170817/GRB~170817A \cite{TheLIGOScientific:2017qsa,GBM:2017lvd,Monitor:2017mdv}. This globally unifying event was viewed in every EM waveband, in GWs, and was searched for in neutrinos as well. 
Gravitational waves enabled astronomers to pin down accurate masses and distance, which are not directly encoded in EM observations, and this was the key to turn an observation of an otherwise faint gamma-ray burst into brand new physics. Combining EM and GW messages immediately determined that one progenitor channel of short gamma-ray bursts comes from neutron star mergers. The speed of gravity and the speed of light were measured to be equal with unprecedented accuracy, and neutron star mergers were found to be the dominant formation mechanism for heavy elements greater then atomic number 43 \cite{Kasen2017}. The richness of these discoveries, from only one event, demonstrates the paradigm-shifting potential of adding the messages encoded in GWs to the information carried by particles and light. 

This one event underlined the necessity for robust, comprehensive data analysis methods to identify, characterize, and distribute rapid GW alerts for EM follow-up. Leadership and communication channels are required for coordinating observations, and early collaboration with EM astronomers will best build the needed infrastructure to maximize the science impact. To encourage and support this collaboration the following efforts are recommended:
\begin{itemize}
\setlength\itemsep{0em}
\item Tight {\bf partnerships with upcoming electromagnetic astronomy missions and surveys} to educate the leadership on both sides about the synergies between GW and EM science.
\item Development of {\bf agreements for alerts and data sharing}. 
\item Support for {\bf a network of small UV/O/IR telescopes} to follow up ground-based alerts, and {\bf radio telescope facilities support} to continue monitoring the pulsar timing array.
\item Development of {\bf public tools and calculators} in collaboration with electromagnetic observers prior to the LISA mission launch.
\item{Universities and GW centers host {\bf faculty sabbaticals} to team up EM and GW experts for long-term collaborations.}
\item{Increase the number of {\bf special sessions and workshops in MMA} at APS and AAS meetings to facilitate interactions between GW and EM experts.}
\item Establishment of a {\bf Center for Multimessenger Astronomy} to provide an interface between GW and EM astronomers, particle astrophysicists, and data scientists. Early efforts of the Scalable Cyberinfrastructure for Multi-messenger Astrophysics \href{https://scimma.org}{SCiMMA} project indicate that the infrastructure, computation, and data science challenges involved in multimessenger astronomy require the focused attention of a center.    
\end{itemize}

A Center for Multimessenger Astronomy will provide the leadership, coordination, and focus to enable discoveries using heterogeneous, massive, sparse, and sometimes proprietary datasets. 
Next-generation survey facilities of every messenger are planning leaps in data collection, inundating us in a  flood of data with volumes and velocities orders of magnitude greater than ever before. 
Doing justice to this opportunity calls for a convergence of theory, computation, observation, instrumentation, statistics, and data science. To facilitate this convergence of global teams we require easily-deployable and flexible platforms for efficient collaboration and communication, as well as significant computing and archiving resources such as what a national center could provide~\cite{Kalogera:2019bdd, scimmawp}.
 

\vspace{10pt} 
\noindent{\large \textmyfont{ Innovations in Low-frequency Gravitational Wave Data Analysis}}

The data analysis challenges for the lower frequency bands of Pulsar Timing and LISA will require expertise from across the fields of astronomy, physics, and data science,  
and present an exciting opportunity for innovative 
data analysis techniques in signal identification, parameter estimation, data access, source catalogs, and multimessenger coordinated observations. 

These innovations 
include general data modeling and analysis, dedicated methods for multimessenger observations, detection of unknown and unmodeled sources, and development of computational and data access tools. 
The ongoing development of data analysis techniques for GW Astronomy faces unsolved problems that are research subjects in their own right. 
These challenges include dealing with data artifacts, 
performing semi-online analysis of LISA data to efficiently update previous source detections and parameter estimations by including the arrival of new data,
building a full statistical inference method to solve a global source separation problem,
and improving and going beyond the time delay interferometry (TDI) algorithm \cite{Tinto2014}, which is the pre-processing method designed to cancel the otherwise overwhelming laser frequency noise.

Unlike the ground-based detections thus far, LISA and PTA sources will be long-lasting, spanning several weeks to the entire mission duration. Where EM studies of LIGO/Virgo discoveries hinge on rapid communication and follow-up observations, low-frequency GW sources present a unique opportunity to provide advanced warning to EM astronomers, with time to coordinate multimessenger observing campaigns of cataclysmic events such as massive black hole mergers out to high redshift. The persistence of the sources present a new data science challenge for GW observers, who will now have to contend with changing behavior of the detectors and many overlapping signals simultaneously present in the data. Like trying to isolate a single conversation at a crowded party, individual GW signals must be separated and characterized to accurately identify and localize the source for further study. New data science techniques are being developed to prevent cross-contamination between overlapping signals, and to develop strategies for optimally interacting with EM observers in this new paradigm. PTA detections of both a background and individual  slowly evolving supermassive black hole binaries over the next decade will pave the way for these  analysis methods. It is critical to provide infrastructure such that lessons learned in one part of the GW spectrum are carried over into the others.

The search for {\it unexpected } GW signals holds the potential for the next great discovery. Understanding the details of the LISA instrument and data stream will be critical in order to identify un-modeled sources. The detection of un-modeled sources with a single LISA detector presents a new challenge beyond Earth-based detectors and pulsar timing arrays. The synthesized LISA data channels are signal-orthogonal, so cross correlation between channels can not be used for detection, and the channels will have correlated noise. 


The community will rely on a {\bf NASA LISA Science Center} 
to provide astronomers with well-documented, continuously-updated, user-friendly tools for data access on multiple levels including community alerts, astrophysical parameter catalogs, open data, and mission guest investigators.
The Gravitational Wave Open Science Center (\href{https://www.gw-openscience.org/about/}{GWOSC}) \cite{GWOSC} is an example started by the LIGO/Virgo collaboration. However, because the potential for revolutionary science is so great, and because the data itself will be of a new type to the astronomical community, a NASA LISA Science Study Team Taskforce report recommends a LISA Science Center on par with a flagship mission, such as Hubble and the Chandra X-ray Center.

 
 For most of the astronomical community, a sophisticated cross-match of GW sources with widely-used astronomical catalogs like Gaia or LSST (i.e., on the basis of position, distance, period, and period change for galactic binaries) is a basic priority to facilitate observing proposals and publications. These new tools need excellent user support, including consultants to assist in customizing pipelines for specific research projects. Some astronomers will need access not only to a catalog of sources, but to the rest of the mission data as well. This is especially important for any claims of deviations from General Relativity or other new physics; such potentially extraordinary discoveries would require strict verification and checks against instrument performance, station-keeping, and early data cleaning, all of which would almost surely require help from staff scientists.

\vspace{10pt} 
\noindent{\large \textmyfont{Technology Development for Gravitational Wave Observatories}}

The technology required for a GW observatory is quite unlike an EM telescope; 
GW observatories, both in space and on the ground, employ laser interferometers to measure any differential shift between test masses. In space, there are free-floating test masses fixed in their local inertial frames and laser interferometers record variations in the light travel time between these test masses, so the goal of the spacecraft is to shield the masses from spurious forces caused by space weather, thermal noise, and electromagnetic radiation. On the ground, likewise, the goal is to minimize environmental coupling into the suspended masses.

Technology teams currently working on GW observatories are small compared to those working in other astronomy fields. This is especially true in the U.S., where only a few ($<$ 5) PI-led groups in academia exist for space-based technology development, in addition to the small teams at NASA and JPL. Ground-based observatories, like LIGO, have slightly more opportunities, though nowhere near the opportunities that currently exist for ground-based EM. 
While this presents a tractable challenge for the development of technology for LISA, {\bf it is a serious problem for future GW observatories.} We recommend the following : 


\begin{itemize}
\vspace{-7pt}
\setlength\itemsep{0em}
\item{{\bf GW technology specific bootcamps}, like the \href{http://www.dunlap.utoronto.ca/training/summer-school/}{Dunlap School} \cite{dunlap}, are an important part of an experimentalist's  training. Hardware techniques that span the spectrum ( i.e. control loops, interferometer noise modeling, electronics, etc) would be especially beneficial. }
\item{Funding for {\bf summer or year-long internships} at GW Instrumentation labs for undergraduates, graduate students and scientists transitioning from other fields into GWs.}
\item{Support for {\bf technology specific conferences and workshops across the GW spectrum}. Currently, there only exists one technology specific conference that brings together technologists across the spectrum (GWADW). Otherwise, technology discussions are held within collaboration-specific meetings.}
\item{ Technology-specific {\bf startup grants to help stimulate building a lab for new faculty hires}. GW faculty ads tend towards theory and data analysis over experimentation due to the additional startup funds needed for hardware.}
\item{{\bf Hardware replicas for trouble-shooting and training} the next generation of experimentalists.}
\item{{\bf Initiating forward-looking programs} that support concepts and technologies beyond those that are directly applicable to the currently planned observatories \cite{apreslisa}}.
\end{itemize}

\vspace{10pt} 
\noindent{\large \textmyfont{ Building an Inclusive field}}


The opportunity to launch a field of physics is a once in a lifetime event; the formation of GW Astronomy has been likened to the birth of Quantum Mechanics in the 1920's. This time, we have the benefit of the science of broadening participation to guide us in best practices for fostering a diverse and inclusive community that values diversity along many axes. 
Our vision is to build a community that is, at the start and at its core, a broad and welcoming one that encourages each of its members to bring their whole-selves into the field, encourages respect and honorable conduct.
We recommend that:
\begin{itemize}
\item{Stakeholders adopt the} \href{https://aas.org/posts/news/2017/02/inclusive-astronomy-nashville-recommendations}{\bf Nashville Recommendations }{\bf \cite{nashville}} from the Inaugural Inclusive Astronomy Meeting
\item{Every GW gathering (conference, workshops, group meetings) has {\bf dedicated time towards learning and practicing inclusive behaviors}. Examples include plenary speakers, codes of conduct, mentoring mixers, discussing equity and inclusion literature, etc.}
\item{Federal funding to hire {\bf Diversity and Inclusion experts to lead regularly-scheduled workshops} for senior, junior and incoming members on inclusive practices and assessments within GW field.} 
\item{{\bf Award} members of the community working to create inclusive GW environments}
\item{Provide funding to {\bf recruit, mentor, and retain} scholars from traditionally underrepresented groups.}
\item{{\bf Climate Evaluations} from Diversity and Inclusion consultants who can provide tailored assessments and recommendations}.
\end{itemize}
 




\vspace{10pt} 
\noindent{\large \textmyfont{ Timeline for GW Astronomy}}

In the case of LISA, there is a sense of urgency that a development program begins now and grows as we approach the launch of the LISA mission in the early 2030s. That urgency is driven both by our previous efforts with building GW analysis and technology tools, and by the LIGO experience — these development efforts only come to fruition on multi-year timescales; some on decade-long timescales. Furthermore, while it will be possible to continue analyzing LISA data after the end of the mission, many (but not all) of the highest priority science goals and most beneficial science results will only be possible during flight, with quasi-real time analysis; a robust analysis technology needs to be in place when observations begin to make maximal science returns. 
Given the scale and complexity of building a new field, a sustained support profile that grows in time, and builds in a scaffolded way, will put the US GW Astronomy community on a trajectory that will deliver robust science.
\vfill

\eject
\bibliography{biblio.bib}

\end{document}